\documentclass[12pt]{article}             
\usepackage{feynmf}                       
\begin{document}
\title{Semiclassical Gauge    \\
           Theories}
\author{Jorge Alfaro\thanks{jalfaro@puc.cl} and Pedro Labra\~{n}a\thanks{plabrana@puc.cl}                                        \\
        Facultad de F\'{\i}sica, Pontificia Universidad Cat\'{o}lica de Chile \\
        Casilla 306, Santiago 22, \textsc{Chile}.
        }
\date{\today}
\maketitle
\begin{abstract}

We study the properties of a non-abelian gauge theory subjected to a gauge invariant
constraint given by the classical equations of motion. The constraint is not imposed by
hand, but appears naturally when we study a particular type of local gauge
transformations. In this way, all standard techniques to treat gauge theories are
available. We will show that this theory lives at one-loop. Also this model retains
some quantum characteristic of the usual non-abelian gauge theories as asymptotic
freedom.
\end{abstract}
%
%
\begin{fmffile}{fmfpat}       
%
%

\section{Introduction}

Yang-Mills Gauge theories have provided the building principle for all the known
interactions of nature. They enable the unified description of the weak and
electro-magnetic interactions as well as the strong interactions, guaranteeing the
renormalizability of the theory.

Gravity is a gauge theory of a different kind. This difference is precisely
the main reason of why we do not have a theory of Quantum Gravity: In Yang-Mills
gauge theories, the gauge symmetry at our disposal enable the formulation of
unitary and renormalizable models. The gauge symmetry of General Relativity, on
the other hand, is not enough to have renormalizability.

The most promising road to quantize gravity is
provided by String Theory \cite{witten}. This
is so because in string theories the amount of gauge symmetries increases enormously.
So much that the model is not only renormalizable, but finite.

One of the motivations of this paper  is to explore new kinds of gauge symmetries
based in the algebraic structure built in the Batalin-Vilkovisky(BV) quantization
method. A preliminary report on these ideas is contained in \cite{Alfaro}.

Some of the theories based on these symmetries share the property of
being one-loop theories (semiclassical), in the sense that higher loop corrections
are absent. So the one-loop effective action is the exact effective
action for these models.As such it could provide a playground to study non-perturbative
effects in Yang-Mills theories.

Another arena where we could apply this program is quantum gravity. t'Hooft and Veltman \cite{thooft} showed that the one loop
effective action of general relativity (pure gravity) is renormalizable on shell. In the semiclassical version of General
Relativity this will be the exact result, so physical observables such as the black hole entropy should be computable for these
models. The computation is too involved in quantum gravity, so we prefer to test our ideas in the simpler case of non-abelian
Yang-Mills theories.

In this paper, we study the properties of a non-abelian gauge theory in which the
expression $\delta(D_{\mu \nu}{F}_{\mu \nu})$ appears in its generating functional.
This constraint is generated, not imposed by hand \cite{gozzi}. To implement it, we
will construct a quantum field model, invariant under a set of local gauge
transformations obtained from (BV) formalism \cite{Alfaro}, then we will show that the
non-abelian gauge theory with constraint is obtained from this model when we study its
generating functional.

We are going to show that it is a one-loop theory . It follows that computations as the $\beta$ function and the effective action
will be exact. Some relevant questions shall be answered: Is it necessary to renormalize the model?, is the theory asymptotically
free?

It will turn out that indeed the theory is renormalizable and
asymptotically free if the gauge group is non-abelian.

The bad news is the presence of negative norm states in the spectrum (see Appendix B),
so the model is not manifestly unitary. It may happens that there could exist a
projection on the Hilbert space such that in the projected subspace the theory is
unitary. In the worst situation , we could consider it as a useful effective theory.

The situation is reminiscent of what happens in Yang-Mills theories on nonsemisimple
groups \cite{T1}. These theories have similar properties as ours. Notice, however, that
our semiclassical models are more general than Yang-Mills theories on nonsemisimple
groups(See equation (\ref{z2})). In particular our models don't need a gauge symmetry to
start with. (See the Appendix A).

The paper is written as follows: In section 2 we introduce the fields, the gauge
transformations and find the invariant Lagrangian that produce the classical equations
of motion as constraints. In section 3, we quantize the model using the BRST formalism
\cite{brst,K} and the background field method \cite{AB}. In section 4, we present the main
result of the paper: the $\beta$ function, and prove that the theory lives at one loop.
In section 5 we analyze the conclusions.

\section{\bf The gauge transformations}

  We work with a compact semisimple Lie group, with
generators $T_{i}$ and structure constants $C^{k}_{ij}$.
 The fields $A^{k}_{\mu}$ and $B^{k}_{\mu}$ have an index $k$ in the algebra
running from $1$ to algebra dimension $N$.
The index $\mu$ goes from $0 \longrightarrow 3$. The space-time metric
is $g_{\mu\nu} = \{-,+,+,+\}$.

We are going to work in the adjoint
representation of the group, then fields and infinitesimal parameters
will be defined as follows $A = A^{k}\, T_{k}$\,, $A_{\mu} = A^{k}_{\mu}\,T_{k}$,
$B = B^{k}\, T_{k}$\,, $B_{\mu} = B^{k}_{\mu}\, T_{k}$\,,
$\varepsilon_{0} = \varepsilon^{i}_{0}\, T_{i}$\,, and
$\varepsilon_{1} = \varepsilon^{i}_{1}\, T_{i}$\,.

Let's continue defining the following transformations obtained from
\cite{Alfaro}. These transformations depend on two parameters and
form a closed algebra. They split into two classes,
homogeneous and inhomogeneous transformations.

The homogeneous transformations are:

\begin{equation}
\label{t1}
\delta_{\varepsilon} A = [\varepsilon_{0},A]\, ,
\end{equation}

\begin{equation}
\label{t2}
\delta_{\varepsilon} B = [\varepsilon_{0},B] + [\epsilon_{1},A].
\end{equation}

 They act over pairs of fields ($A$, $B$), which can be thought as components
of one field $\Phi$, where its first and second  component transforms according
to (\ref{t1}) and (\ref{t2})respectively.

The inhomogeneous transformations are:

\begin{equation}
\label{t3}
\delta_{\varepsilon} A_{\mu} = \partial_{\mu} \varepsilon_{0} +
[\varepsilon_{0},A_{\mu}] \, ,
\end{equation}

\begin{equation}
\label{t4}
\delta_{\varepsilon} B_{\mu} = [\varepsilon_{0},B_{\mu}] +
          \partial_{\mu} \varepsilon_{1} + [\varepsilon_{1},A_{\mu}].
\end{equation}

  They act over pairs of fields ($A_{\mu}$, $B_{\mu}$) ,which can
be thought as one field $\Psi_{\mu} = (A_{\mu}, B_{\mu})$.

 We can see that under a transformation with $\varepsilon_{0}$ parameter
fields $A$ and $B$ transform as scalars in the adjoint representation of
the group, and $A_{\mu}$ transforms as a usual gauge field.
Under $\varepsilon_{1}$ transformation $A$ and $A_{\mu}$ don't transform
while $B$ and $B_{\mu}$ transform mixing with their partners.

The covariant derivatives act over $\Phi$-type fields and
are defined as a function of the transformation as follows:

\begin{equation}
\label{t5a}
D_{\mu} A = \partial_{\mu} A - \left[ A_{\mu}, A \right]\, ,
\end{equation}

\begin{equation}
\label{t5b}
D_{\mu} B = \partial_{\mu} B - \left[ A_{\mu}, B \right] -
                               \left[ B_{\mu}, A \right],
\end{equation}
and transform homogeneously as components of a
$\Phi$-type field.

The field strength is obtained, as usual, commuting
two covariant derivatives. We get:

\begin{equation}
\label{dem3}
\left[ D_{\mu},D_{\nu} \right]
\left( \begin{array}{c}
                   A \\
                   B \\
             \end{array} \right) =
\left( \begin{array}{c}
     \left[A,F_{\mu \nu} \right] \\
  \left[B,F_{\mu \nu} \right] + \left[A,\tilde{F}_{\mu \nu} \right] \\
             \end{array} \right),
\end{equation}

where we have defined:

\begin{equation}
\label{dem4}
 \begin{array}{c}
 F_{\mu \nu} = \partial_{\mu}A_{\nu} -
               \partial_{\nu}A_{\mu} + [A_{\nu},A_{\mu}]\, ,  \\
                                                          \\
\tilde{F}_{\mu \nu} =  \partial_{\mu}B_{\nu} - \partial_{\nu}B_{\mu} +
                      [B_{\nu},A_{\mu}] + [A_{\nu},B_{\mu}].\\
             \end{array}
\end{equation}

 They transform homogeneously under a gauge transformation and can be grouped
into a $\Phi$-type field.

  If we work with compact semisimple Lie groups \cite{Cap21},
we can normalize $T_i$ as follows:$Tr \left( T_{i}\, T_{j} \right) = -2 \delta_{ij}$.

Since the field strength transform homogeneously we have the following gauge invariant Lagrangian:

\begin{equation}
\label{L3}
{\cal L} = -\frac{1}{4g^{2}} F^{k}_{\mu \nu}F^{k}_{\mu \nu} +
\frac{1}{g'}F^{k}_{\mu \nu}\tilde{F}^{k}_{\mu \nu}\, ,
\end{equation}
with

\begin{equation}
\label{F}
\begin{array}{c}
F^{k}_{\mu \nu} = \partial_{\mu}A^{k}_{\nu} -
\partial_{\nu}A^{k}_{\mu} +
C^{k}_{ij}\, A^{i}_{\nu}\, A^{j}_{\mu}\, ,                  \\
                                                        \\
\tilde{F}^{k}_{\mu \nu} = \partial_{\mu}B^{k}_{\nu} -
\partial_{\nu}B^{k}_{\mu}
+ C^{k}_{ij}\, B^{i}_{\nu}\, A^{j}_{\mu} +
C^{k}_{ij}\, A^{i}_{\nu}\, B^{j}_{\mu} \, ,                     \\
             \end{array}
\end{equation}

 The first term in (\ref{L3}) is a usual gauge invariant
Lagrangian density while the second term is a new expression linear in
the field $B_{\mu}$.

The constant $g$ and $g'$ can be absorbed by the tensors
$F_{\mu \nu}$ and $\tilde{F}_{\mu \nu}$ if we do a change of
scale; $A_{\mu} \longrightarrow g A_{\mu}$,
$B_{\mu} \longrightarrow \tilde{g} B_{\mu}$ and $g' = g\tilde{g}$.
We can note that the theory has only one coupling constant $g$.

It is interesting to notice that $A_{\mu}$ satisfies the classical Yang-Mills
equations of motion and transform as a usual gauge field whereas
$B_{\mu}$ will generate a constraint in the path integral.


 Let us consider the generating functional for the system:

\[
Z = \int\! {\cal D}A_{\mu}\, {\cal D}B_{\mu}\,  e^{iS}\, ,
\]
with

\[
S = \int\! d^{4}\! x \left\{
-\frac{1}{4g^{2}} F^{k}_{\mu \nu}F^{k}_{\mu \nu} +
\frac{1}{g'}F^{k}_{\mu \nu}\tilde{F}^{k}_{\mu \nu} \right\}.
\]

Now, we integrate over $B_{\mu}$ to get:
\[
Z = \int\! {\cal D}A_{\mu}\, {\cal D}B_{\mu}\,  e^{iS_{0}}\,
e^{-i \frac{2}{g'}\int\! d^{4}\! x\,
\left( D_{\mu} F_{\mu \nu} \right)^{k} B^{k}_{\nu}}
\]
\begin{equation}
\label{delta}
= \int\! {\cal D}A_{\mu}\, e^{iS_{0}}\,
\delta\! \left( D_{\mu} F_{\mu \nu} \right)^{k}\, ,
\end{equation}
with

\[
S_{0} = -\frac{1}{4g^{2}} \int\! d^{4}\! x \:\,
F^{k}_{\mu \nu}F^{k}_{\mu \nu}\, .
\]

Thus we see that the model is equivalent to a non-abelian gauge theory subjected to the constraint. Taking advantage of this
situation we will study the gauge theory with the constraint studying this new model (\ref{L3}) by placing $A_{\mu}$ and $B_{\mu}$
on the same level, forgetting for a moment that $A_{\mu}$ satisfies the classical equation of motion and that $B_{\mu}$ is an
auxiliary field. The kind of diagrams that we can construct will show us that $B_{\mu}$ is an auxiliary field and that this model
corresponds to a theory of the same kind as (\ref{delta}).

From the expression (\ref{delta}) we can note that if we integrate
$A_{\mu}$ we obtain:

\begin{equation}
\label{z1}
Z = {det}^{-1} {\left| \frac{\delta^{2} S}{\delta A_{\mu}^{2}} \right|}_{0}
e^{-i\, S_{0}}\, ,
\end{equation}
where both, the determinant and $S_{0}$, are evaluated in the classical
solution of $A_{\mu}$.
This expression, times a factor $2$, is the semiclassical value
of $Z$ . This is a hint that the theory lives at one loop.

It's interesting to notice that the new term in the Lagrangian could be
generated in the following way.
First we calculate the variation of ${\cal L}_{0}$ respect to $A_{\mu}$:

\begin{equation}
\label{z2}
\delta {\cal L}_{0} =
\delta\! \left( F^{k}_{\mu \nu}\,F^{k}_{\mu \nu} \right) =
\frac{\delta {\cal L}_{0}}{\delta A^{l}_{\alpha}}\,\delta A^{l}_{\beta} +
\frac{\delta {\cal L}_{0}}
{\delta\! \left( {\partial}_{\beta}A^{l}_{\alpha}\right)}\,\delta\!
\left(\partial_{\beta}A^{l}_{\alpha}\right),
\end{equation}
then, we define $B_{\mu} = \delta A_{\mu}$ as a new independent field.
 The new expression that becomes from (\ref{z2}) it's:

\[
{\cal L}' = 2\,F^{k}_{\mu \nu}\,\tilde{F}^{k}_{\mu \nu}\, ,
\]
where $F^{k}_{\mu \nu}$ and $\tilde{F}^{k}_{\mu \nu}$ were defined in
(\ref{dem4}).

The transformation (\ref{t4}) could be found in a similar way
if we calculate the variation of $\delta A_{\mu}$ and
defining $\delta \varepsilon_{0} = \varepsilon_{1}$ as a new parameter.

\[
\delta B_{\mu} = \delta \left( \delta A_{\mu} \right)
= \delta \left( \partial_{\mu} \varepsilon_{0} +
[\varepsilon_{0},A_{\mu}] \right)
\]
\[
= \partial_{\mu} \varepsilon_{1} + [\varepsilon_{1},A_{\mu}] +
[\varepsilon_{0},B_{\mu}] .
\]

In this way we can recover our previous results.This way of proceeding is general and
it could be used in any Lagrangian with similar results, even if there isn't a gauge
symmetry to start with. (See the Appendix A).



\section{\bf The Model at the Quantum Level}

  We will quantize our model (\ref{L3}) using the BRST formalism \cite{brst,K}
and the background field method \cite{AB}. The BRST transformations are:

\begin{equation}
\label{brst5}
\left\{
\begin{array}{l}
\delta' A^{k}_{\mu} = -\left( \partial_{\mu}\eta^{k}_{0} +
                           C^{k}_{ij}\, \eta^{i}_{0}A^{j}_{\mu} \right) \\
                                                                        \\
\delta' B^{k}_{\mu} = -\left( \partial_{\mu}\eta^{k}_{1} +
                   C^{k}_{ij}\,  \eta^{i}_{1}A^{j}_{\mu} +
                        C^{k}_{ij}\,  \eta^{i}_{0}B^{j}_{\mu} \right)\, , \\
\end{array}
\right.
\end{equation}


\begin{equation}
\label{brst6}
\left\{
\begin{array}{l}
\delta '\eta^{k}_{0} = \frac{1}{2} C^{k}_{ij}\, \eta^{i}_{0}\eta^{j}_{0} \\
                                                                          \\
\delta '\eta^{k}_{1} = C^{k}_{ij}\, \eta^{i}_{0}\eta^{j}_{1}\, ,          \\
\end{array}
\right.
\end{equation}


\begin{equation}
\label{brst7}
\left\{
\begin{array}{lll}
\delta '\bar{C}_{0} = ib_{0}  & &   \delta 'b_{0} = 0    \\
                                                          \\
\delta '\bar{C}_{1} = ib_{1}  & &   \delta 'b_{1} = 0 \, .                                        \\
\end{array}
\right.
\end{equation}

Let us write $A_{\mu} \rightarrow A_{\mu} + \cal{A}_{\mu}$ and
$B_{\mu} \rightarrow B_{\mu} + \cal{B}_{\mu}$, where
$(A_{\mu},B_{\mu})$ are the background fields and
$(\cal{A}_{\mu}, \cal{B}_{\mu})$ are the quantum fields.
Then gauge fields strength (\ref{F}) decompose as
follow:

\begin{equation}
\label{nota}
\begin{array}{c}
F^{k}_{\mu \nu} = {\bf F}^{k}_{\mu \nu} +
\left( D_{\mu}\cal{A}_{\nu}\right)^{k} - \left( D_{\nu}\cal{A}_{\mu}\right)^{k}
- C^{k}_{ln}\,{\cal{A}}^{l}_{\mu}\,{\cal{A}}^{n}_{\nu} \\
                                                        \\
\tilde{F}^{k}_{\mu \nu} = {\bf \tilde{F}}^{k}_{\mu \nu} +
\left( D_{\mu}\cal{B}_{\nu}\right)^{k} - \left( D_{\nu}\cal{B}_{\mu}\right)^{k}
- C^{k}_{ln} \left( {\cal{A}}^{l}_{\mu}\,{\cal{B}}^{n}_{\nu} -
                       {\cal{A}}^{l}_{\nu}\,{\cal{B}}^{n}_{\mu} \right)\,  \\
\end{array}
\end{equation}
where ${\bf F}_{\mu \nu}$ and ${\bf \tilde{F}}_{\mu \nu}$ are the fields
strength evaluated on the background fields and the covariant derivatives
only contain $(A_{\mu},B_{\mu})$ fields.


Now, as usual in the BRST method \cite{K}, we add a null operator $-i\,\delta'(P)$
to the original Lagrangian and use it as a gauge fixing and FP-ghost.
In our case we use
$P = \frac{2}{g'}\bar{C}^{k}_{0}\left( D_{\mu}\cal{A}_{\mu} \right)^{k} +
\bar{C}^{k}_{1}\left(D_{\nu}\cal{B}_{\nu} \right)^{k} -
\bar{C}^{k}_{0}b^{k}_{1}$.

The functional integral over the auxiliary fields ($b_{0}$, $b_{1}$) can be done
obtaining the following effective Lagrangian:

\begin{equation}
\label{bf15}
{\cal L} = -\frac{1}{4g^{2}}F^{k}_{\mu \nu}F^{k}_{\mu \nu} +
\frac{1}{g'}F^{k}_{\mu \nu}\tilde{F}^{k}_{\mu \nu} +
\frac{2}{g'}
\left( D_{\mu}\cal{B}_{\mu} \right)^{k}\!
\left(D_{\nu}\cal{A}_{\nu} \right)^{k} +
{\cal L}_{FP}\, ,
\end{equation}
where the explicit forms of the fields strength are (\ref{nota}) and

\begin{equation}
\label{fp}
\begin{array}{c}
{\cal L}_{FP} =
\frac{2i}{g'}\bar{C}^{k}_{0}
\left[\partial_{\mu}\partial_{\mu} \delta^{ki} +
C^{k}_{ij}\, \vec \partial_{\mu}\left(A^{j}_{\mu} + {\cal A}^{j}_{\mu}\right) +
      C^{k}_{ij}\, A^{j}_{\mu}\vec \partial_{\mu} \right.                        \\
                                                                           \\
\left.
+\, C^{k}_{\alpha j}\, C^{\alpha}_{i \beta}\left( A^{\beta}_{\mu} +
{\cal A}^{\beta}_{\mu} \right)A^{j}_{\mu} \right] \eta^{i}_{0}       \\
                                                                       \\
+\, i\bar{C}^{k}_{1}
\left[ \partial_{\mu}\partial_{\mu} \delta^{ki} +
 C^{k}_{ij}\, \vec \partial_{\mu} \left(A^{j}_{\mu} + {\cal A}^{j}_{\mu} \right)
 + C^{k}_{ij}\, A^{j}_{\mu}\vec \partial_{\mu}  \right.                         \\
                                                                             \\
\left.
+\, C^{k}_{\alpha j}\, C^{\alpha}_{i \beta} \left( A^{\beta}_{\mu} +
        {\cal A}^{\beta}_{\mu} \right) A^{j}_{\mu} \right] \eta^{i}_{1} \\
                                                                         \\
+\, i\bar{C}^{k}_{1}
\left[ C^{k}_{ij}\,\vec \partial_{\mu}
\left( B^{j}_{\mu} + {\cal B}^{j}_{\mu} \right) +
C^{k}_{\alpha j}\, C^{\alpha}_{i \beta} \left( B^{\beta}_{\mu} +
 {\cal B}^{\beta}_{\mu} \right)A^{j}_{\mu}
+ C^{k}_{ij}\, B^{j}_{\mu} \vec \partial_{\mu}  \right.                     \\
                                                                         \\
\left.
+\, C^{k}_{\alpha j}\, C^{\alpha}_{i \beta} \left( A^{\beta}_{\mu} +
            {\cal A}^{\beta}_{\mu} \right)B^{j}_{\mu}
            \right] \eta^{i}_{0}\, .                                         \\
\end{array}
\end{equation}

It follows that the Feynman rules are \cite{AB,TH}:
{\bf Propagators:}

\begin{equation}
{\cal A}^{k}_{\mu} \:
\parbox{20mm}{\begin{fmfgraph}(20,15)
     \fmfleft{i} \fmfright{o}
     \fmf{photon}{i,o}
      \end{fmfgraph}} \:{\cal B}^{s}_{\nu}
 = \frac{i}{(2 \pi)^{4}}\frac{g'}{2}\delta^{ks}
                            \frac{\delta_{\mu \nu}}{p^{2}} \,\, ,
\end{equation}
\begin{equation}
{\cal B}^{k}_{\mu} \:
\parbox{20mm}{\begin{fmfgraph}(20,15)
     \fmfleft{i} \fmfright{o}
     \fmf{curly}{i,o}
      \end{fmfgraph}} \: {\cal B}^{s}_{\nu}
= \frac{i}{(2 \pi)^{4}}
\left(\frac{g'}{g} \right)^{2} \frac{\delta^{ks}}{4} \frac{1}{p^{2}}
 \left[ \delta_{\mu \nu} - \frac{p_{\mu}p{\nu}}{p^{2}} \right] \, ,
\end{equation}

\begin{equation}
\bar{C}^{i}_{0} \:
\parbox{20mm}{\begin{fmfgraph}(20,15)
     \fmfleft{i} \fmfright{o}
     \fmf{dots_arrow}{i,o}
      \end{fmfgraph}} \: \eta^{j}_{0}
= -\frac{1}{(2 \pi)^{4}}\left(\frac{g'}{2}\right)\frac{\delta^{ij}}{p^{2}}  \,\, ,
\end{equation}

\begin{equation}
\bar{C}^{i}_{1} \:
\parbox{20mm}{\begin{fmfgraph}(20,15)
     \fmfleft{i} \fmfright{o}
     \fmf{dashes_arrow}{i,o}
      \end{fmfgraph}} \: \eta^{j}_{1}
= -\frac{1}{(2 \pi)^{4}}\frac{\delta^{ij}}{p^{2}}  \,\, .
\end{equation}
%
{\bf Vertex:}

We write explicitly the vertices we are going to use in the calculation of
the $\beta$ function. Additional vertices can be read off directly from (\ref{bf15}).

a) Gauge Fields:
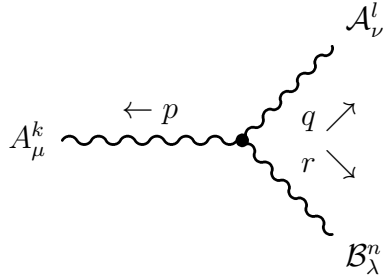
\begin{center}
\begin{fmfchar*}(40,25)
  \fmfleft{em} \fmflabel{$A^{k}_{\mu}$}{em}
\fmf{photon,label=$\leftarrow p$}{Z,em}
\fmf{photon, label=$r \searrow$}{fb,Z}
\fmf{photon, label=$q \nearrow$}{f,Z}
  \fmfright{fb,f} \fmflabel{${\cal B}^{n}_{\lambda}$}{fb}
  \fmflabel{${\cal A}^{l}_{\nu}$}{f}
  \fmfdot{Z}
\end{fmfchar*}
\end{center}

\begin{equation}
\Gamma^{kln}_{\mu \nu \lambda} (p,q,r)
= -(2 \pi)^{4}\left( \frac{2}{g'}C^{n}_{kl} \right)
\left[\delta_{\mu \lambda}(r + q - p)_{\nu} +
\delta_{\nu \lambda}(q - r)_{\mu} + \delta_{\mu \nu}(p - q -r)_{\lambda}\right]  \, .
\end{equation}

b) Ghosts:

i)
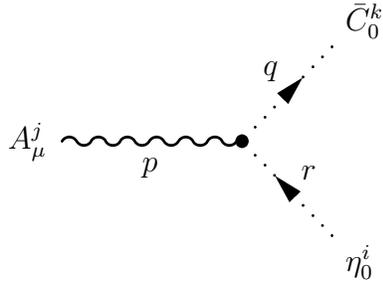
\begin{center}
\begin{fmfchar*}(40,25)
  \fmfleft{em} \fmflabel{$A^{j}_{\mu}$}{em}
\fmf{photon,label=$p$}{em,Z}
\fmf{dots_arrow, label=$r$}{fb,Z}
\fmf{dots_arrow, label=$q$}{Z,f}
  \fmfright{fb,f} \fmflabel{$\eta^{i}_{0}$}{fb}
  \fmflabel{$\bar{C}^{k}_{0}$}{f}
  \fmfdot{Z}
\end{fmfchar*}
\end{center}

\begin{equation}
\Gamma^{ijk}_{\mu}(q,r) = (2 \pi)^{4}i\left(\frac{2}{g'}\right)
                              C^{k}_{ij}[q + r]_{\mu}  \,\, ,
\end{equation}


ii)
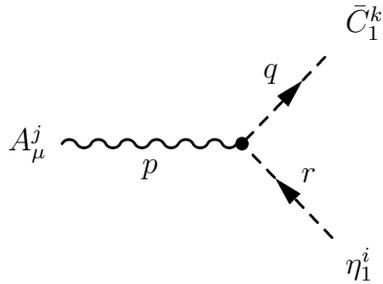
\begin{center}
\begin{fmfchar*}(40,25)
  \fmfleft{em} \fmflabel{$A^{j}_{\mu}$}{em}
\fmf{photon,label=$p$}{em,Z}
\fmf{dashes_arrow, label=$r$}{fb,Z}
\fmf{dashes_arrow, label=$q$}{Z,f}
  \fmfright{fb,f} \fmflabel{$\eta^{i}_{1}$}{fb}
  \fmflabel{$\bar{C}^{k}_{1}$}{f}
  \fmfdot{Z}
\end{fmfchar*}
\end{center}

\begin{equation}
\Gamma^{ijk}_{\mu}(q,r) = (2 \pi)^{4}i
                              C^{k}_{ij}[q + r]_{\mu}  \,\, .
\end{equation}


\section{The $\beta$ Function}

Let's  proceed with the computation of the $\beta$ function as a first
way to explore the properties of the model.
The diagrams that contribute to this are listed below:


\begin{equation}
A^{k}_{\mu}\:
\parbox{20mm}{\begin{fmfgraph}(20,15)
\fmfleft{i} \fmfright{j}
\fmflabel{$A^{k}_{\mu}$}{i}
\fmf{photon}{i,v1} \fmf{photon}{j,v2}
\fmf{photon, left, tension=.3}{v1,v2}
\fmf{photon, left, tension=.3}{v2,v1}
\fmfdot{v1,v2}
\end{fmfgraph}}  \: A^{k'}_{\mu'} =
i\frac{C^{n}_{kl}\, C^{n}_{k'l}}{(4 \pi)^{2}}\, \frac{20}{3}\,
(p^{2}\delta_{\mu \mu'} - p_{\mu}p_{\mu'})\frac{1}{\epsilon}\, ,
\end{equation}

\begin{equation}
\parbox{20mm}{\begin{fmfchar*}(20,15)
\fmfleft{i} \fmfright{j}
\fmflabel{$A^{j}_{\mu}$}{i}
\fmf{photon}{i,v1} \fmf{photon}{j,v2}
\fmf{dots, left, tension=.3}{v1,v2}
\fmf{dots, left, tension=.3}{v2,v1}
\fmfdot{v1,v2}
\end{fmfchar*}} \: A^{j'}_{\mu'} =
(-1)\, i\,\frac{C^{k}_{ij}\,C^{k}_{ij'}}{(4  \pi)^{2}}\, \frac{1}{3}\,
(p_{\mu}p_{\mu'} - p^{2}\delta_{\mu \mu'})\frac{1}{\epsilon}\, ,
\end{equation}

\begin{equation}
\parbox{20mm}{\begin{fmfchar*}(20,15)
\fmfleft{i} \fmfright{j}
\fmflabel{$A^{j}_{\mu}$}{i}
\fmf{photon}{i,v1} \fmf{photon}{j,v2}
\fmf{dashes, left, tension=.3}{v1,v2}
\fmf{dashes, left, tension=.3}{v2,v1}
\fmfdot{v1,v2}
\end{fmfchar*}} \: A^{j'}_{\mu'} =
(-1)\, i\, \frac{C^{k}_{ij}\, C^{k}_{ij'}}{(4 \pi)^{2}}\, \frac{1}{3}\,
(p_{\mu}p_{\mu'} - p^{2}\delta_{\mu \mu'})\frac{1}{\epsilon}\, ,
\end{equation}
where we have written only the pole part of the graphs.

 Adding the three diagrams, we get:

\[
A^{k}_{\mu}
\parbox{20mm}{\begin{fmfgraph}(20,15)
\fmfleft{i} \fmfright{j}
\fmf{photon}{i,v1} \fmf{photon}{j,v2}
\fmf{photon, left, tension=.3}{v1,v2}
\fmf{photon, left, tension=.3}{v2,v1}
\fmfdot{v1,v2}
\end{fmfgraph}}A^{k'}_{\mu'} +
A^{k}_{\mu}
\parbox{20mm}{\begin{fmfgraph}(20,15)
\fmfleft{i} \fmfright{j}
\fmf{photon}{i,v1} \fmf{photon}{j,v2}
\fmf{dots, left, tension=.3}{v1,v2}
\fmf{dots, left, tension=.3}{v2,v1}
\fmfdot{v1,v2}
\end{fmfgraph}}A^{k'}_{\mu'} +
A^{k}_{\mu}
\parbox{20mm}{\begin{fmfchar*}(20,15)
\fmfleft{i} \fmfright{j}
\fmf{photon}{i,v1} \fmf{photon}{j,v2}
\fmf{dashes, left, tension=.3}{v1,v2}
\fmf{dashes, left, tension=.3}{v2,v1}
\fmfdot{v1,v2}
\end{fmfchar*}}A^{k'}_{\mu'}
\]
\begin{equation}
\label{d2}
= i\, \frac{C^{n}_{kl}\,C^{n}_{k'l}}{(4 \pi)^{2}}\, \frac{22}{3}\,
(p^{2}\delta_{\mu \mu'} - p_{\mu}p_{\mu'})\frac{1}{\epsilon}\, .
\end{equation}

   Since explicit gauge invariance is retained in the background field
method \cite{AB}, we have the following relation between the $\beta$ function
and the field renormalization factor $Z_{A}$:

\begin{equation}
\label{cap46}
\beta = -g \left[ \beta_{0} \left( \frac{g}{4 \pi} \right)^{2} +
\beta_{1} \left( \frac{g}{4 \pi} \right)^{4} \right] ,
\end{equation}

\begin{equation}
\label{cap47}
Z_{A} = 1 + \frac{1}{\epsilon} \beta_{0} \left( \frac{g}{4 \pi} \right)^{2} +
\frac{1}{2 \epsilon} \beta_{1} \left( \frac{g}{4 \pi} \right)^{4} \, .
\end{equation}


Then, our $\beta$ function is:

\begin{equation}
\label{beta}
\beta = - \frac{g^{3}}{16 \pi^{2}} C_{2} \frac{22}{3}\,\, ,
\end{equation}
where $C_{2}$  is the Quadratic Casimir operator.

We see that the theory is asymptotically free.


Now we will show that the model get corrections from one loop contributions only.

  Let's begin the proof with diagrams that contain only fields
${\cal A}_{\mu}$ and ${\cal B}_{\mu}$ propagating in them. To form these diagrams we have the propagator $P_{1}$ that connects field
${\cal A}_{\mu}$ with ${\cal B}_{\mu}$, the propagator $P_{2}$ that connects ${\cal B}_{\nu}$
with itself and the vertices type
$V_{1}$ and type $V_{2}$. Vertices $V_{1}$ only contain one leg ${\cal B}$ and its other legs are $A$ or ${\cal A}$. Vertices
$V_{2}$ only contain $A$ or ${\cal A}$ legs.

  The quantum fields live only inside loops \cite{AB}, and we don't have
a vertex with two legs ${\cal B}$, then the following condition must be
satisfied to form a diagram:

\begin{equation}
\label{con}
P_{1} + 2P_{2} = V_{1}.
\end{equation}

The number of loops in a diagram is

\begin{equation}
\label{L}
L = P - V + 1 \, ,
\end{equation}
if (\ref{con}) is satisfied, the equation (\ref{L}) becomes

\begin{equation}
L = 1 - V_{2}\, .
\end{equation}

Therefore it's impossible to form diagrams with $L > 1$ that contain only fields ${\cal A}$ and ${\cal B}$ in them.

  Let's continue with diagrams that contain ghost field. The ghost's
propagators connect $\bar{C}_{0}$ with $\eta_{0}$ and $\bar{C}_{1}$ with
$\eta_{1}$

 The forms of the vertices are

\begin{equation}
\begin{array}{cc}
\left( \bar{C}_{0}\, \eta_{0}\, {\cal A} ... \right) \; \,
                \left( \bar{C}_{1}\, \eta_{0}\, B ... \right)\\
                                                                \\
    \left( \bar{C}_{1}\, \eta_{1}\, {\cal A} ... \right) \; \,
              \left( \bar{C}_{1}\, \eta_{0}\, {\cal B} ... \right) \\
    \end{array}
\end{equation}

   To construct diagrams with $L > 1$ that contain ghost field inside,
we must include $P_{1}$ propagator. Therefore we have
to include the vertex $( \bar{C}_{1}\, \eta_{0}\, {\cal B} ... )$
in this diagrams, but we can't connect this vertex with the other vertices
keeping the ghost field inside the loop, because the vertex
$( \bar{C}_{0}\, \eta_{1} ... )$ doesn't exist.

 Therefore it's impossible to form diagrams with $L > 1$.

 Also it's a important comment that all external legs of our diagrams are
ending in $A$ fields, then quantum correction to the effective action only depend on $A$. This corrections have been generated by
the new term in the Lagrangian, because  all the propagators and vertex that we used  depend on this term.


\section{Conclusions}

  In this paper, we have studied the properties of a non-abelian gauge theory
subjected to a  gauge invariant constraint given by the classical equations
of motion. For this purpose we constructed a model of quantum
field theory from a type of local gauge transformations and proved that this
model is equivalent to the gauge theory subjected to a constraint.

 We proved that the theory lives at one-loop and that the quantum corrections
to the effective action depend only on the usual gauge field $A_{\mu}$. Also we can
note that the scattering amplitudes with external legs $A_{\mu}$ only exist at the tree
level and the loop correction gives non-trivial scattering amplitudes only with
$B_{\mu}$ as external legs. Such amplitudes are absent at the tree level. This type of
results are similar to those obtained in \cite{T1}, although our motivation and
starting point were entirely different.

  We computed the $\beta$ function and from this result
we conclude that the theory is asymptotically free, retaining in this way the usual non-abelian gauge theory properties. Moreover
because the theory lives at one loop  the computations of the $\beta$ function is exact.

Having tested our ideas in the simpler case of Yang-Mills theories, we would like to
apply the prescription to General Relativity. We expect to find a one loop theory of
quantum gravity, which is renormalizable on shell. The most important issue in this
context is the lack of manifest unitarity in semiclassical gauge theories, due to the
presence of negative norm states in the spectrum(See Appendix B). It may be possible to
truncate the Hilbert space so that in the restricted space the theory is both unitary
and renormalizable.

{\bf Acknowledgments}. We thank A. Andrianov for useful discussions.
The work of JA  is partially supported by Fondecyt 1010967. The work
of PL is partially supported by a CONICYT Fellowship.


%
\end{fmffile}
%

%
{\bf Appendix A:}

We begin with a Lagrangian ${\cal L}_{0}$ that depends on a field $\phi$.
Then we calculate the variation of ${\cal L}_{0}$ respect to $\phi$ and
define ${\cal L}'$ as follows:

\begin{equation}
\label{CG}
{\cal L}' =
\frac{\delta {\cal L}_{0}}{\delta \phi}\,\delta \phi +
\frac{\delta {\cal L}_{0}}
{\delta\! \left( {\partial}_{\mu}\phi \right)}\,\delta\!
\left(\partial_{\mu}\phi \right) =
\frac{\delta {\cal L}_{0}}{\delta \phi}\,\tilde{ \phi} +
\frac{\delta {\cal L}_{0}}
{\delta\! \left( {\partial}_{\mu}\phi \right)}\,
\partial_{\mu}\tilde{\phi}\, ,
\end{equation}
were we have defined $\tilde{\phi} = \delta \phi$ as a new independent
field. The new action is:

\[
S = \int\! d^{4}\! x \left\{{\cal L}_{0} + {\cal L}'  \right\} =
\int\! d^{4}\! x \left\{ {\cal L}_{0} +
\left( \frac{\delta {\cal L}_{0}}{\delta \phi} -
\partial_{\mu}\! \left[ \frac{\delta {\cal L}_{0}}
{\delta\! \left( {\partial}_{\mu}\phi \right)}\right]
\right) \tilde{\phi}\right\}.
\]

Then the generating functional will be:

\[
Z = \int\! {\cal D}\phi\, {\cal D}\tilde{\phi}\,  e^{iS}\, =
\int\! {\cal D}\phi\,\, e^{iS_{0}}\,\delta(Classical\, Eq.\, for\, \phi)
\]

The theories constructed following this scheme have properties similar
to our model, for instance, they live at one-loop level and only the
field that satisfies the classical equation of motion appears in the
quantum corrections to the effective action.

{\bf Appendix B:}

\title{Semiclassical Klein-Gordon Field}
\maketitle

\vspace{0.5 cm}

Let us study the unitarity problem in a simpler model than Semiclassical Yang-Mills
Theories, but constructed following the same scheme.

The new Lagrangian obtained from the free Klein-Gordon Lagrangian is:

\begin{equation}
\label{ap1}
{\cal L} = \frac{1}{2}{\partial}_{\mu} \phi\, {\partial}^{\mu} \phi
- \frac{1}{2}\,m^2\, \phi^2 + {\partial}_{\mu} \phi\, {\partial}^{\mu} {\bar\phi}
- m^2\,\phi\,{\bar\phi}  \, ,
\end{equation}

The Hamiltonian is:

\begin{equation}
\label{ap4}
H = \int\! d^{3} x\,
\left(
\pi\, {\bar \pi} - \frac{1}{2}\, {\bar \pi}^2 +
\frac{1}{2}\left(\nabla \phi \right)^2 +
\nabla \phi \cdot  \nabla {\bar \phi} +
\frac{1}{2}\,m^2\, \phi^2 + m^2\, \phi\,{\bar \phi}\, \right)  \, ,
\end{equation}
where we have defined the conjugate momenta in the usual way

\begin{equation}
\label{ap2}
\pi(x) = \frac{\partial {\cal L}}{\partial {\dot \phi}(x)} =
{\dot \phi}(x) + {\dot {\bar \phi}}(x)
\end{equation}

\begin{equation}
\label{ap3}
{\bar \pi}(x) = \frac{\partial {\cal L}}{\partial {\dot {\bar \phi}}(x)} =
{\dot \phi}(x)
\end{equation}

We write the solution of the classical equation of motion as Fourier
modes:\footnote{Four-vectors are denoted by light italic type and three-vectors are
denoted by boldface type.}

\begin{equation}
\label{phi}
\phi({\bf x},t) = \int\! \frac{d^{3}p}{(2\pi)^3}\,\frac{1}{\sqrt{2 E_{p}}}
\left.
\left[ a_{\bf p}e^{-ip \cdot x} +
a^{\dag}_{\bf p}e^{ip \cdot x }
\right]\right|_{p_{0} = E_{p}}
\end{equation}

\begin{equation}
\label{phib}
{\bar \phi}({\bf x},t) = \int\! \frac{d^{3}p}{(2\pi)^3}\,\frac{1}{\sqrt{2 E_{p}}}
\left.
\left[ {\bar a}_{\bf p}e^{-ip \cdot x} +
{\bar a}^{\dag}_{\bf p}e^{i p \cdot x}
\right]\right|_{p_{0} = E_{p}}
\end{equation}
with,

\[
E_{p} = + \sqrt{|{\bf p}|^2 + m^2} .
\]

To quantize the model we impose the usual canonical commutation relations. The only
non-vanishing commutators are:

\begin{equation}
\label{co}
\left[a_{\bf p},{\bar a}^{\dag}_{\bf p'}\right] =
\left[{\bar a}_{\bf p},a^{\dag}_{\bf p'}\right] =
- \left[{\bar a}_{\bf p},{\bar a}^{\dag}_{\bf p'}\right] =
(2\pi)^{3}\,\delta^{(3)}({\bf p} - {\bf p'})
\end{equation}

In terms of ladder operators, the Hamiltonian is:

\begin{equation}
\label{H}
H =  \int\! \frac{d^{3}p}{(2\pi)^3}\,E_{p} \left(
a^{\dag}_{\bf p}a_{\bf p} + a^{\dag}_{\bf p}{\bar a}_{\bf p} +
{\bar a}^{\dag}_{\bf p}a_{\bf p} -\, {\bar a}^{\dag}_{\bf p}{\bar a}_{\bf p}
\right) ,
\end{equation}
where we have neglected the infinite constant term.

The state $| 0 \rangle$ such that $a_{\bf p}\, | 0 \rangle = 0$ and
${\bar a}_{\bf p}\, | 0 \rangle  = 0$
for all ${\bf p}$ is the ground state. All other energy eigenstates
can be built by acting on $| 0 \rangle$ with creator operator
$a^{\dag}_{\bf p}$
or ${\bar a}^{\dag}_{\bf p}$. In general the state $a^{\dag}_{\bf p}\cdots
{\bar a}^{\dag}_{\bf p'}\cdots | 0 \rangle$ is an eigenstate with energy
$E_{p} + E_{p'} + \cdots$.

Note that the state of the form ${\bar a}^{\dag}_{\bf p}| 0 \rangle$ have
negative norm while the states of the form  $a^{\dag}_{\bf p}| 0 \rangle$
have null norm.

Let us define the following operators:

\begin{equation}
\label{def1}
b_{\bf p} = a_{\bf p} + {\bar a}_{\bf p} \hspace{1cm}b^{\dag}_{\bf p} = a^{\dag}_{\bf p} + {\bar a}^{\dag}_{\bf p}
\end{equation}

The algebra of these operators is

\begin{equation}
\label{def3}
\left[b_{\bf p},b^{\dag}_{\bf p'}\right] =
- \left[{\bar a}_{\bf p},{\bar a}^{\dag}_{\bf p'}\right] =
(2\pi)^{3}\,\delta^{(3)}({\bf p} - {\bf p'})
\end{equation}

\begin{equation}
\label{co1}
\left[b_{\bf p},b_{\bf p'}\right] =
\left[b_{\bf p},b^{\dag}_{\bf p'}\right] =
\left[b_{\bf p},{\bar a}_{\bf p'}\right] =
\left[b_{\bf p},{\bar a^{\dag}}_{\bf p'}\right] =
\left[{\bar a}_{\bf p},{\bar a}_{\bf p'}\right] =
\left[{\bar a^{\dag}}_{\bf p},{\bar a}^{\dag}_{\bf p'}\right] = 0
\end{equation}

We write the Hamiltonian in term of $b_{\bf p}$, $b^{\dag}_{\bf p}$,
${\bar a^{\dag}}_{\bf p}$, and  ${\bar a}_{\bf p}$ obtaining:

\begin{equation}
\label{H1}
H = \int\! \frac{d^{3}p}{(2\pi)^3}\,E_{p} \left[
2b^{\dag}_{\bf p}b_{\bf p} - {\bar a}^{\dag}_{\bf p}{\bar a}_{\bf p}
\right] .
\end{equation}

We can note that $b^{\dag}_{\bf p}$ create a eigenstate of $H$ with energy $2E_{p}$ and
positive norm. The states created with $b^{\dag}_{\bf p}$ have zero inner product with
the states created using ${\bar a}^{\dag}_{\bf p}$, then we can divide the $H$
eigenstates into two orthogonal subspaces. The subspace generated by $b^{\dag}_{\bf p}$
has no problem with negative norm and have a positive spectrum.

This situation is reminiscent of negative norm modes in QED and it may be possible to
find a unitary truncation of the Hilbert space that produces a theory free of any
degrees of freedom with negative probability.
\end{document}